\documentclass[aps,prl,amsmath,amssymb,superscriptaddress,showkeys,showpacs,twocolumn,floatfix]{revtex4-1}% Physical Review Letters=
\usepackage[final]{graphicx}
\usepackage{hyperref}
\usepackage{amsmath}
\usepackage{bbm}
\usepackage{amsfonts}
\usepackage{amssymb}
\usepackage{latexsym}
\usepackage{graphicx}
\usepackage[english]{babel}
\usepackage{multirow}
\usepackage{float}
\usepackage{url}
\usepackage{slashed}
\usepackage{xcolor}
\usepackage[utf8]{inputenc}
\usepackage{verbatim}
\usepackage{lipsum}

\usepackage[utf8]{inputenc}

\usepackage{diagbox}

\usepackage{mathtools}
\usepackage{amsfonts}
\usepackage{mathrsfs}
\usepackage{bbm}
\usepackage{slashed}

\usepackage{graphicx}
\usepackage{color}
\usepackage{array}
\usepackage{esint}
\usepackage{placeins}
\usepackage{booktabs}
\usepackage{makecell}
\usepackage{epstopdf}
\usepackage[caption=false]{subfig}

\usepackage{xspace}
\usepackage{siunitx}
\usepackage{hyperref}
\usepackage[nameinlink]{cleveref}
\usepackage{appendix}

\usepackage{xifthen}
\usepackage{xcolor}
\hypersetup{
	colorlinks,
	linkcolor={red!75!black},
	citecolor={blue!75!black},
	urlcolor={blue!75!black}
}
\usepackage{comment}

\newcommand{\be}{\begin{equation}}
\newcommand{\ee}{\end{equation}}
\newcommand{\ba}{\begin{array}}
\newcommand{\ea}{\end{array}}

\usepackage{ulem}
\newcommand{\beq}{\begin{eqnarray}}
\newcommand{\eeq}{\end{eqnarray}}

\newcommand{\ii}{\mathrm{i}}

\begin{document}

\title{Baryogenesis Induced by Magnetic Field Effects During the Electroweak Phase Transition}
\author{Yuefeng Di}
\affiliation{
Institute of Theoretical Physics, 
Chinese Academy of Sciences, Beijing 100190, China}
\affiliation{School of Physical Sciences, University of Chinese Academy of Sciences (UCAS), Beijing 100049, China}

\author{Ligong Bian\thanks{Corresponding Author.}}
\email{lgbycl@cqu.edu.cn}
\affiliation{Department of Physics and Chongqing Key Laboratory for Strongly Coupled Physics, Chongqing University, Chongqing 401331, P. R. China}
\author{Rong-Gen Cai}
\email{cairg@itp.ac.cn}
\affiliation{
Institute of Fundamental Physics and Quantum Technology, Ningbo University, Ningbo, 315211, China}

\date{\today}

\begin{abstract}
We numerically investigate the first-order electroweak phase transition in the background of a hypermagnetic field with three-dimensional lattice simulation. The generation of baryon asymmetry is observed, and we present the relationship between baryon number asymmetry and magnetic field strength and its helicity. 
We find the magnetic field strength required to achieve the correct matter-antimatter asymmetry is about $10^{-17}\sim10^{-14}$ Gauss at present, depending on the correlation length of the helical magnetic field. This study provides a mechanism for explaining the baryon number asymmetry with cosmic magnetic fields. 

\end{abstract}
\maketitle
%\tableofcontents

\noindent{\it Introduction.-}
Magnetic fields (MFs) are ubiquitous in various astrophysical environments in the Universe, including the solar system, the Milky Way~\cite{Wielebinski2005}, galaxies, galaxy clusters~\cite{Clarke_2001, Bonafede_2010, Feretti_2012}, and even the voids of large-scale structures~\cite{Dolag_2010, Tavecchio_2010, Tavecchio_2011, Vovk_2012}. These MFs are thought to originate from primordial MFs generated during some physical processes in the early Universe, such as the { first-order electroweak phase transition (PT) motivated by new physics beyond the Standard Model (SM) of particle physics }~\cite{PhysRevLett.51.1488, Di_2021, Enqvist:1993np,Yang_2022,Vachaspati:1991nm,Baym:1995fk,Grasso:1997nx}, the {first-order} QCD PT~\cite{Quashnock:1988vs}, and inflation~\cite{Turner:1987bw, Martin:2007ue, Ratra:1991bn, Subramanian_2009, Kunze_2010, Adshead:2016iae, Kanno:2009ei}, followed by amplification mechanisms~\cite{Grasso_2001,Widrow_2002,Kulsrud_2008,Kandus_2011, Widrow_2011,Durrer_2013}. {These primordial MFs might affect the dark matter prediction of axion~\cite{Campanelli:2005ye}, the Big Bang nucleosynthesis~\cite{Luo:2018nth}, and the spectrum of the cosmic microwave background~\cite{Mack:2001gc,Subramanian:2015lua,Jedamzik:1999bm,Sethi:2004pe,Saga:2019ora}.}  The potential new physics responsible for seeding those large-scale MFs{, such as the duration of inflation~\cite{Kahniashvili:2016bkp}, the PT scale~\cite{Durrer_2013},} can be investigated through gamma-ray observations of blazars~\cite{Dermer_2011, Taylor_2011, Neronov_2010}, {and detection of stochastic gravitational wave backgrounds~\cite{Caldwell:2022qsj,Roshan:2024qnv}.}

Previous studies have demonstrated that the presence of MFs at zero temperature influences the Higgs vacuum, inducing a vortex structure known as the Ambj\o rn-Olesen condensation~\cite{Nielsen:1978rm, PhysRevD.45.3833, SALAM1975203, LINDE1976435,Ambjorn_1989, Chernodub_2023}.  MFs can affect both the baryon number violation process during the electroweak cross-over~\cite{Annala:2023jvr,Ho:2020ltr} and the first-order PT by altering the electroweak sphaleron energy~\cite{Comelli:1999gt} and PT dynamics~\cite{Kajantie:1998rz}. In addition, the connection between electroweak strings and baryon number asymmetry generation has been established through the relationship between the Chern-Simons number and the MF helicity associated with the Z field~\cite{PhysRevLett.101.171302, PhysRevLett.87.251302,Vachaspati:1994ng,Barriola:1994ez}.

Crucially, assuming the occurrence of a first-order electroweak PT~\cite{Hindmarsh_2021, Caldwell:2022qsj, Athron_2024}, the Chern-Simons number stored in the MF is proposed to convert into baryon number via chiral anomaly, thereby generating the baryon number asymmetry of the Universe (BAU)~\cite{Giovannini_1998, Giovannini:1997eg, Joyce:1997uy, Fujita:2016igl}. Such a PT can create the necessary out-of-equilibrium conditions to explain the BAU puzzle through the electroweak baryogenesis mechanism~\cite{Morrissey_2012, Cohen_2012}. Furthermore, the stochastic gravitational wave backgrounds associated with the strong first-order  PT~\cite{Mazumdar_2019, Caprini_2016, Caprini_2020} can be detected by future space-based gravitational wave observatories, such as the Laser Interferometer Space Antenna (LISA)~\cite{LISA:2017pwj}, TianQin~\cite{Luo_2016}, and Taiji~\cite{taiji}.

In this Letter, we investigate the first-order electroweak PT in a {preexisting} 
MF background using lattice simulations. We analyze the impact of MFs on the PT, establish for the first time the quantitative relationship between baryon number asymmetry and MF strength, and determine the present MF strength required to achieve the correct matter-antimatter asymmetry without relying on electroweak baryogenesis. For generality, we consider both helical MFs with infinite correlation length and with characteristic correlation lengths associated with the electroweak PT.

This Letter employs natural units where $\hbar=c=k_\mathrm{B}=1$. Given the rapid completion of the first-order electroweak PT, we neglect the effect of the expansion of the universe, the evolution of background  MFs, and temperature changes.

\noindent{\it The Simulation Framework.-}
%\label{sec:Model}
We consider the Lagrangian of the electroweak theory extended with a background  U(1) field $Y_\mathrm{ex}^\mu$ whose field strength $Y_{\mathrm{ex}}^{\mu\nu}$ remains constant in time:  
\begin{align}
    \mathcal{L} =\ &(D_\mu\Phi)^\dagger(D^\mu\Phi)
    - \frac14 W^a_{\mu\nu}W^{a\mu\nu}
    - \frac14 Y_{\mu\nu}Y^{\mu\nu}\notag\\ 
    &\ - \frac{1}{2} Y^{\mathrm{ex}}_{\mu\nu} Y^{\mu\nu} - \mathcal{V} (\Phi)\;.\label{L}
\end{align}
Here, $\Phi$ denotes the Higgs doublet, while $W^a_{\mu\nu}$ and $Y_{\mu\nu}$ represent the field strengths of the $\mathrm{SU(2)_L}$ and $\mathrm{U(1)_Y}$ gauge fields, respectively. 
The covariant derivative is given by
\begin{align}
    D_\mu = \partial_\mu - \ii g\frac{\sigma^a}{2} W_\mu^a - 
    \ii g' \,\frac12(Y_\mu + Y^\mathrm{ex}_\mu), \label{D}
\end{align}
where $\sigma^a,\ a=1,\ 2,\ 3$ are Pauli matrices, and the coupling constants are $g = 0.65$ and $g' = 0.53g$. The fields 
$W^\mu$ and $Y^\mu = (Y^0,\ \boldsymbol{Y})$ correspond to the  
$\mathrm{SU(2)_L}$ and $\mathrm{U(1)_Y}$ gauge fields, respectively. The external U(1) field $Y^\mathrm{ex}_{\mu}$, which generates the time-independent external MF, is explicitly included, while the term  $Y^\mathrm{ex}_{\mu\nu} Y^{\mu\nu}_{\mathrm{ex}}$ is omitted as it contributes only a constant. For the specific implementation of the discrete model, we refer to the {\it Equations of Motion on Lattice} and {\it Initialization} sections in the {\it supplemental material}.

The potential $\mathcal{V} (\Phi)$ contains a potential barrier and admits a first-order electroweak PT{, as suggested by new physics beyond the SM~\cite{Hindmarsh_2021, Caldwell:2022qsj, Athron_2024},} given by
\begin{align}
    \mathcal{V} (\Phi) &= -\mu^2  \Phi^\dagger\Phi + A(\Phi^\dagger\Phi)^{3/2} + \lambda (\Phi^\dagger\Phi)^2\;,\label{V2}
\end{align}
{
which indicates the cubic Higgs coupling deviation from the SM of particle physics, and can be probed by high-energy colliders in the future~\cite{FCC:2018byv,Arkani-Hamed:2015vfh}.}
Using the \texttt{FindBounce}~\cite{Guada_2020}, we adopt the initial bubble profile
\begin{align}
    \Phi(r) &= \frac v2\left[1-\tanh\left(\frac{r-R_0}{l_\mathrm{w}}\right)\right]
    \begin{pmatrix}
        0 \\ 1
    \end{pmatrix} ,\label{bubbleprofile} \\
    \dot\Phi(r) &= 0,
\end{align}
where $v$ is the vacuum expectation value of the Higgs field, $r$ denotes the distance from bubble center, $R_0$ denotes the radius of bubble, and $l_\mathrm{w}$ denotes the bubble wall thickness. In the simulation, each lattice site in the false vacuum at each time step has a probability $p_\mathrm{bubble}$ to nucleate a vacuum bubble. When a bubble is nucleated, it will expand and collide with another to propel the PT process. 

%For infinite correlation length hyperMF, we assume that it is a time-invariant homogeneous hyperMF 
%along the $z$ direction $\boldsymbol{B}_Y^\mathrm{ex} = (0,\ 0,\ B_Y^\mathrm{ex})$ with
%the corresponding external U(1) gauge field being $Y^{\mathrm{ex}\mu} = (0,\ 0,\ xB_Y^\mathrm{ex},\ 0)$. This is the setup of the MF without helicity. 

%To achieve first order PT, 
%We use the following type of the potential function $\mathcal{V} $ with $(\Phi^\dagger\Phi)^{3/2}$, 
%\begin{align}
 %   \mathcal{V} (\Phi) &= -\mu^2  \Phi^\dagger\Phi + A(\Phi^\dagger\Phi)^{3/2} + \lambda (\Phi^\dagger\Phi)^2, \label{V2}
%\end{align}
%. 

According to the chiral anomaly in electroweak theory, the relationship between fermions and gauge fields is described by \cite{Adler:1969gk, Bell:1969ts, tHooft:1976rip, tHooft:1976snw}:
\begin{align}
    \partial_\mu j_\mathrm{B}^\mu = N_\mathrm{g}\left[\frac{g^2}{16\pi^2}\mathrm{Tr}\,(W_{\mu\nu} \tilde W^{\mu\nu}) - \frac{g'^2}{32\pi^2}Y_{\mu\nu} \tilde Y^{\mu\nu}\right], 
\end{align}
where $N_\mathrm{g} = 3$ denotes the number of fermion generations, and $\tilde X^{\mu\nu}=\frac12\varepsilon^{\mu\nu\rho\sigma} X_{\rho\sigma}$ is the Hodge dual. Integrating this equation over a finite time interval and infinite volume yields
\begin{align}\label{eq:nbs}
    n_\mathrm{B} = N_\mathrm{g}\frac{\Delta N_\mathrm{CS}(t)}{V}
    = N_\mathrm{g}\frac{N_\mathrm{CS}(t)-N_\mathrm{CS}(0)}{V},
\end{align}
where the Chern-Simons number $N_\mathrm{CS}(t)$ is given by \cite{Moore_2002}
%\begin{widetext}
\begin{align}
    \frac{N_\mathrm{CS}(t)}{V} 
    &=\frac1V \frac{1}{32\pi^2}\varepsilon^{ijk}\int \mathrm{d}^3\boldsymbol{x} \biggl [ 
        -g'^2 (Y_i+Y^\mathrm{ex}_i) (Y_{jk}+Y^\mathrm{ex}_{jk})\nonumber\\
 &\ \ \ \ +g^2 \left ( W^a_i W^a_{jk}  -\frac{g}{3} \varepsilon^{abc} W^a_i W^b_j W^c_k \right )
   \biggr ]\label{eq:ncs}
\end{align}
%\end{widetext}
and $V=L^3$ is the lattice volume. The U(1) component of $N_\mathrm{CS}$ corresponds to the hypermagnetic helicity: 
\begin{align}
    h_{Y} = \frac{H_{Y}}{V} &= \frac1V \varepsilon^{ijk}\int \mathrm{d}^3\boldsymbol{x}\,
    (Y_i+Y_i^\mathrm{ex}) (Y_{jk}+Y_{jk}^\mathrm{ex}) \notag\\ 
    &= \frac1V \int \ \mathrm{d}^3\boldsymbol{x}\,
    (\boldsymbol{Y}+\boldsymbol{Y}^\mathrm{ex})
    \cdot (\boldsymbol{B}_Y+\boldsymbol{B}^\mathrm{ex}_Y),
\end{align}
up to a negative proportionality factor.

{During the first-order PT, bubble collisions induce a dramatic change of the Higgs field $\Phi$ in the $g'\mathrm{Im}[\Phi^\dagger D_i\Phi]$ term of the equation of motion about $Y_i$ (see the \textit{Equations of Motion on Lattice} of the supplemental material for details), causing $Y_i$ to undergo drastic changes, and results in} non-zero contributions from $\boldsymbol{Y}\cdot\boldsymbol{B}_Y$, $\boldsymbol{Y}\cdot\boldsymbol{B}^\mathrm{ex}_Y$ and  $\boldsymbol{Y}^\mathrm{ex}\cdot\boldsymbol{B}_Y$ in $N_\mathrm{CS}$ and $h_\mathrm{Y}$. 
{The U(1)$_\mathrm{Y}$ gauge field strength $Y_{\mu\nu} = \partial_\mu Y_\nu - \partial_\nu Y_\mu$ also contributes to the electromagnetic field in the form of $
    A_{\mu\nu} \supset  Y_{\mu\nu} \cos\theta_W$~\cite{Di_2021}. } 
%electromagnetic field strength since which includes the gradient part of the Higgs field ~\cite{tHooft:1974kcl,Vachaspati:1991nm}:
%$
   % A_{\mu\nu} \supset  - \ii \frac{2}{gv^2}[(D_\mu\Phi)^\dagger (D_\nu\Phi) - (D_\nu\Phi)^\dagger (D_\mu\Phi)]\sin\theta_W$,
%that generate MF~\cite{Di_2021}, and results in} non-zero contributions from $\boldsymbol{Y}\cdot\boldsymbol{B}_Y$, $\boldsymbol{Y}\cdot\boldsymbol{B}^\mathrm{ex}_Y$ and  $\boldsymbol{Y}^\mathrm{ex}\cdot\boldsymbol{B}_Y$ . 
%{When bubbles collide, the Higgs field $\Phi$ between the two bubbles will undergo drastic changes, which will affect the $g'\mathrm{Im}[\Phi^\dagger D_i\Phi]$ term in the equation of motion about $Y_i$, causing $Y_i$ to undergo drastic changes. $Y_i$ also } 
In other words, the time variation of the {Chern-Simons number and} hypermagnetic helicity arises from the combined effects of the {preexisting} external helical hyperMF and the MF generated during the first-order PT.

The baryon asymmetry of the Universe can be calculated through the relationship between the change in baryon number and the variation of the Chern-Simons number: 
\begin{align}
    \eta_\mathrm{B}(t) = \frac{n_\mathrm{B}(t)}{s}=\frac{45N_\mathrm{g}}{2\pi^2 g_{*S}T^3}\frac{\Delta N_\mathrm{CS}(t)}{V} \;.\label{eq:etaB}
    %= 0.064\frac{1}{T^3}\frac{\Delta N_\mathrm{CS}(t)}{V}.
\end{align}
Here, $g_{*S}=106.75$ is the effective number of degrees of freedom in entropy before the PT. 

%The physical quantity that measures the net baryon number density in the universe is $\eta_\mathrm{B} = n_\mathrm{B}/s = (8.6\pm0.1)\times10^{-11}$ \cite{refId0, PhysRevD.98.030001}, which is a constant if particles are neither produced nor destroyed. $s = 2\pi^2 g_{*S}T^3/45$ is the entropy density. $g_{*S}$ is the effective number of degrees of freedom in entropy, which is equal to $106.75$ before the electroweak PT. 

%When the helicity of the external hyperMF is not $0$, that is, $\boldsymbol{Y}^\mathrm{ex}\cdot\boldsymbol{B}^\mathrm{ex}_Y\neq0$, 

%Moreover, the greater the strength and helicity of the external hyperMF, the greater the increase in the latter two terms, and the more drastic and large the change will be. 

For generality, in this Letter, we consider two types of hyperMF background with distinct correlation lengths, defined as 
$\lambda_{BY} = \int \mathrm{d}k\, k^{-1}E_{BY}(k)\ \Big/ \int \mathrm{d}k\, E_{BY}(k)$, 
where $\int \mathrm{d}k\, E_{BY}(k) = \int \mathrm{d}^3\boldsymbol{x}\, \frac12 [\boldsymbol{B}^\mathrm{ex}_Y(\boldsymbol{x})]^2/V$. For simplicity, we model an infinite correlation length hyperMF with helicity by setting $Y^{\mathrm{ex}\mu} = (0,\ 0,\ xB_Y^\mathrm{ex},\ h_\mathrm{factor}LB_Y^\mathrm{ex})$, where $h_\mathrm{factor}=0,\pm1$, and $L = N\Delta x$ represents the edge length of the lattice box. 
For hyperMFs with finite correlation lengths, we adopt the formulation from~\cite{Brandenburg_2020}:
$   \Tilde{B}_{Yi}^\mathrm{ex}(\boldsymbol{k}) = B_\mathrm{ini}\Theta(k-k_\mathrm{UV})\left(\delta_{ij}-\hat{k}_i\hat{k}_j-\ii\sigma_\mathrm{M}\varepsilon_{ijl}\hat{k}_l\right)g_j(\boldsymbol{k})k^n$. 
Here $\Theta(k-k_\mathrm{UV})$ is the Heaviside function, ensuring the hyperMF has a finite correlation length comparable to the mean bubble separation $R_*=(V/N_\mathrm{bubble})^{1/3}$ during the percolation process of vacuum bubbles in the PT. The parameter $k_\mathrm{UV}$ denotes the ultraviolet cutoff wave number, and $\boldsymbol{g}(\boldsymbol{k})$ is the Fourier transform of a Gaussian-distributed random vector field that is $\delta$-correlated in all three dimensions, i.e., $g_i(\boldsymbol{x})g_j(\boldsymbol{x}')=\delta_{ij}\delta^3(\boldsymbol{x-x'})$. The spectral index $n$ determines the power-law behavior of the spectrum, while the helicity degree is controlled by $\sigma_\mathrm{M}=0,1,-1$, corresponding to non-helical, maximally positive, and maximally negative helicities, respectively. 
The specific implementation of hyperMFs on the lattice is detailed in the {\it Hypermagnetic Field on Lattice} section of the {\it supplemental material}. 

We conduct numerical simulations to investigate the scenario of hyperMFs with infinite correlation lengths on a three-dimensional lattice of size $N^3 = 128^3$, employing lattice spacing $\Delta x$ and periodic boundary conditions for real-time evolution. For each set of hyperMFs or helicity configurations, we perform 20 runs with fixed parameters and average the results. 
%And, to investigate the effects of the correlation length, we study the case of $\lambda\sim R_0$ (and $R_*$) with $N=128$ (and $N=512$). 
To explore the effects of finite correlation lengths on the first-order electroweak PT, we perform simulations on two lattice sizes, $N = 128$ and $N=512$, with correlation lengths $\lambda_B\sim R_*/4\sim R_0$ and $\lambda_B\sim R_*$, respectively, where $R_0$ is the initial vacuum bubble radius.  We examine four different hyperMF strengths by varying the spectral index $n$, with each strength including three helicity cases ($\sigma_\mathrm{M}=0,\pm1$). 
When $k_\mathrm{UV} = 2\Delta k$, {the resulting value of (comoving) correlation lengths for different n ($n=0,2,2,3$) are approximately equal. }Due to computational constraints, for the $N=512$ case, we limit the simulations to 4 runs for each combination of $\sigma=\pm 1$  and hyperMF strength. {The parameters that remain unchanged in the simulation are presented in Tab. \ref{tab:parameter}.}
\begin{table}[!htp]
    \centering
    \begin{tabular}{c|c|c|c|c|c|c|c|c}
        \hline\hline
        $v$ &$\Delta x$& $\Delta t$&$\mu^2$ & $A$ &$\lambda$ & $R_0$ & $l_\mathrm{w}$ & $p_\mathrm{bubble}$ \\
        \hline
        $1$ &0.2&0.04& 0.68 & 2.59 & 1.60 &$13\Delta x$ &$6\Delta x$&$5\times10^{-8}$ \\
        \hline\hline
    \end{tabular}
    \caption{Parameters setup in the simulation. }
    \label{tab:parameter}
\end{table}

%\section{Numerical Results} \label{sec:NR}
%The time interval $\Delta t = \Delta x/5$. 
%We fix $p_\mathrm{bubble}=5\times10^{-8}$, $R_0=16\Delta x$ and $l_\mathrm{w}=6\Delta x$.

%when the MF increases beyond the first critical value $g'B_{Y\mathrm{c1}}=eB_\mathrm{c1}=m_W^2$ where $m_W$ is the mass of the W boson, in order to stabilize the vacuum, the Higgs field will form a hexagonally arranged vortex, 

%which is called  

%\begin{figure}[!htp]
 %   \centering
  %  \includegraphics[scale=0.08]{phi2nb20.png}
   % \includegraphics[scale=0.075]{Phi2tzAvehel20.png}
    %\caption{$\overline {\Phi^2}$ in the case with non-helical (Left) and helical (Right) homogeneous hyperMF $g'B_Y^\mathrm{ex}/m_W^2=3.63$ at $t/R_*\simeq35$. }
    %\label{fig:condens}
%\end{figure}

\begin{figure}[!htp]
    \centering
    \includegraphics[scale=0.5]{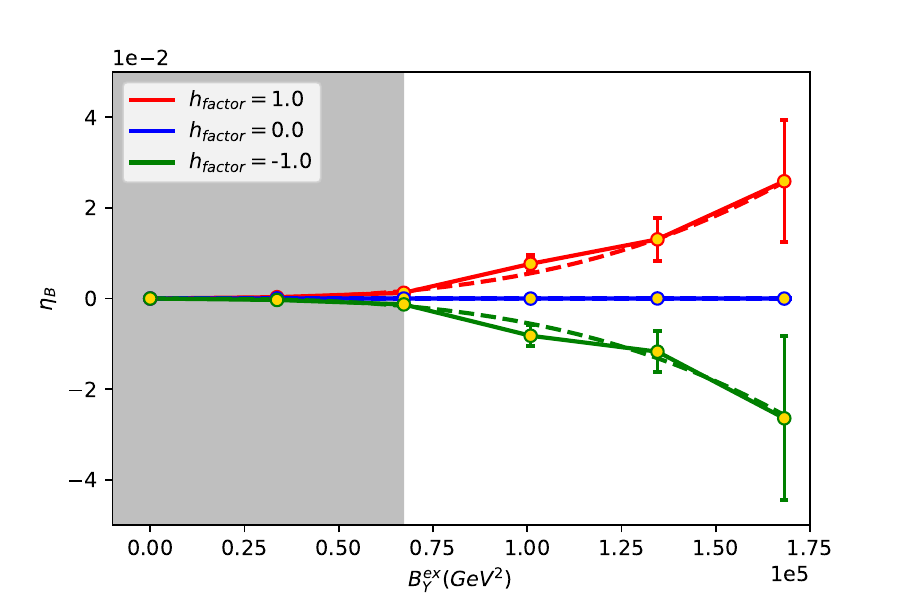}
    \caption{$\eta_\mathrm{B}$ changes with helical homogeneous hyperMF strength whose $h_\mathrm{factor}=\pm1,0$. The dashed line is the fitting result of Eq.~\eqref{eq:fit}. The gray area indicates the magnetic field strength where Higgs condensation has not yet occurred. The error bar
    indicates that $95\%$ of the data falls on the bar.} 
    \label{fig:Ncshel}
\end{figure}

\noindent{\it Numerical Results.-}We now investigate the generation of baryon asymmetry during the first-order PT. Fig.~\ref{fig:Ncshel} illustrates the variation of $\eta_\mathrm{B} $ with respect to the hyperMF strength for the case of an external hyperMF with infinite correlation length. Fig.~\ref{fig:Ncshel} demonstrates two things. Firstly, as the hyperMF strength increases, $\eta_\mathrm{B}$ also increases, indicating enhanced baryon production. Secondly, the sign of $\eta_\mathrm{B}$ depends on the sign of $h_\mathrm{factor}$, and if the external hyper MF is non-helical, there will be no baryon production. 
The data points in Fig.~\ref{fig:Ncshel} are fitted with the dashed line, yielding the following relationship:
\begin{align}
    \eta_B = 5.5\times10^{-6}h_\mathrm{factor}\left(\frac{B_Y^\mathrm{ex}}{10^4\,\mathrm{GeV}^2}\right)^3. \label{eq:fit}
\end{align}

%The left panel of Fig.~\ref{fig:helvsCS} shows the change of the stabilized $\Delta n_\mathrm{CS}$ under different $h_{Y0}$, while the right panel shows the change of $\eta_\mathrm{B}$ over time under different $h_{Y0}$. 

\begin{figure}[!htp]
    \centering
    \includegraphics[scale=0.5]{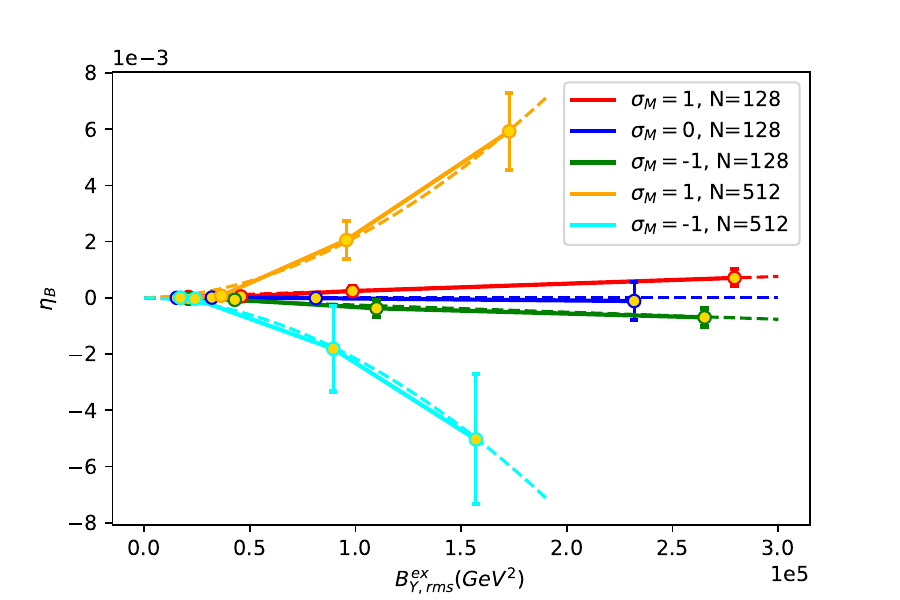}
    \caption{The variation of $\eta_\mathrm{B}$ with different hyperMF strength and correlation length ($N=128$ corresponds to $\lambda_B \sim R_0$, $N=512$ corresponds to $\lambda_B \sim R_*$.). The dashed lines are the fitting results of Eq.~\eqref{eq:fitspec} and Eq.~\eqref{eq:fitspec512}, respectively. The root mean square hyperMF stength is $B_{Y,\mathrm{rms}}^\mathrm{ex} = \sqrt{2\rho_{BY}^\mathrm{ex}}$ with $\rho_{BY}^\mathrm{ex}$ being the volume averaged hyperMF energy density.%Bottom: The evolution of $\eta_\mathrm{B}$ over time under different $\sigma_\mathrm{M}$. The spectral index is fixed to $n=0$. 
    }
    \label{fig:SBCS}
\end{figure}

We now turn to the scenario of a hyperMF with a finite correlation length comparable to the mean bubble separation. Numerical simulations confirm that as the MF energy increases, $\eta_\mathrm{B}$ for $\sigma_\mathrm{M} = \pm 1$ deviates significantly from zero, while $\eta_\mathrm{B}$ for $\sigma_\mathrm{M} = 0$ remains close to zero due to the absence of helical MF effects. 
The results are shown in Fig. \ref{fig:SBCS}, and the dashed lines correspond to the following fitting formulas: 
\begin{align}
    \eta_\mathrm{B} &= 2.5\times10^{-5}\sigma_\mathrm{M}\left(\frac{B_{Y\mathrm{,rms}}^\mathrm{ex}}{10^4 \,\mathrm{GeV}^2}\right)\;,\;\lambda_B\sim R_0\;.\label{eq:fitspec} \\
    \eta_\mathrm{B} &= 3.0\times10^{-5}\sigma_\mathrm{M}\left(\frac{B_{Y\mathrm{,rms}}^\mathrm{ex}}{10^4 \,\mathrm{GeV}^2}\right)^{13/7}\;,\;\lambda_B\sim R_*\;.\label{eq:fitspec512} 
\end{align}

Based on the fitting formulas provided in Eqs.~(\ref{eq:fit}, \ref{eq:fitspec}, \ref{eq:fitspec512}), we summarize the relationship between the baryon number density and the strength and correlation length of the external hyperMF as follows: 
\begin{align}
    \eta_\mathrm{B}\sim 10^{-5}\sigma_\mathrm{M}\left(\frac{B_{Y\mathrm{,rms}}^\mathrm{ex}}{10^4 \,\mathrm{GeV}^2}\right)^{\frac{12\lambda_B+R_*}{4\lambda_B+3R_*}}. \label{eq:etaVSlambda}
\end{align}
{
 We observe that the correlation length of the hyperMF influences the power-law relationship between $\eta_\mathrm{B}$ and $B_Y^\mathrm{ex}$. More exactly, a large mean bubble separation $R_*$ for hyperMF with finite correlation length yields a large baryon asymmetry generation. The generation of $\eta_\mathrm{B}$ is driven by the variation of the Chern-Simons number $N_\mathrm{CS}$ during the PT. Details on that and the accompanied electroweak sphaleron rate can be found in the accompanied article~\cite{Di:2025ncl}.}

\begin{figure}[!htp]
    \centering
    \includegraphics[scale=0.4]{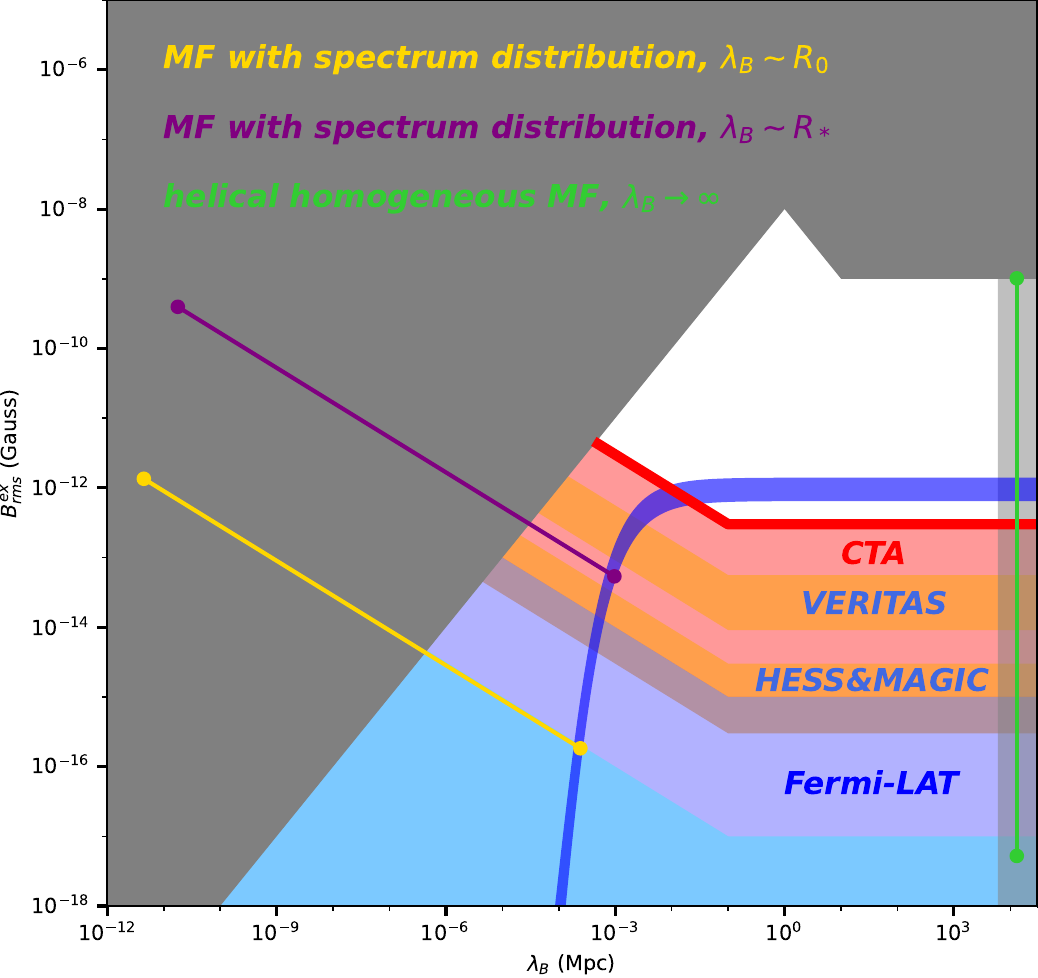}
    \caption{The yellow line, purple line, and green line show the evolution of strength and correlation length of MFs that can yield the correct baryon asymmetry. The blue band represents the range of MF strength and correlation length required to achieve correct baryon asymmetry based on Eq. \eqref{eq:etaVSlambda}. The light gray region on the right represents the Hubble scale. The gray region is excluded by MHD turbulent decay and CMB anisotropy measurements~\cite{Durrer_2013}. The projected sensitivity of the Cherenkov Telescope Array observations to intergalactic MF strength is marked with a red region~\cite{CTA:2020hii}. Orange regions are excluded at the 95\% to 99\% confidence level from H.E.S.S.~\cite{HESS:2014kkl}, MAGIC~\cite{MAGIC:2010goh} and VERITAS~\cite{VERITAS:2017gkr}. The dark and light blue regions are excluded by the lower bounds on IGMF strength from Fermi-LAT~\cite{Taylor_2011, Fermi-LAT:2018jdy}.}
    \label{fig:blazer}
\end{figure}

\noindent{\it Cosmological Implications.-}For a homogeneous MF, regardless of helicity, its evolution follows the scaling law:  $B_\mathrm{rec}/B_\mathrm{EW} = (\tau_\mathrm{rec}/\tau_\mathrm{EW})^{-5/7}$~\cite{Brandenburg_Durrer_Huang_Kahniashvili_Mandal_Mukohyama_2020}, where $\tau$ denotes conformal time, and the subscripts ``rec" and ``EW" refer to the recombination epoch and the electroweak PT epoch, respectively. After recombination, the MF evolves adiabatically with the comoving MF strength and correlation length remain unchanged. Combining this with  Eq.~\eqref{eq:fit}, we find that the homogeneous helical MF capable of generating the correct baryon asymmetry corresponds to a present-day field strength of approximately  $5.2\times10^{-18}$ Gauss. 

For MFs with finite correlation lengths, under fully helical conditions, the comoving correlation scale evolves as $\lambda_\mathrm{rec}/\lambda_\mathrm{EW} = (\tau_\mathrm{rec}/\tau_\mathrm{EW})^{2/3}$, and the comoving MF strength evolves as $B_\mathrm{rec}/B_\mathrm{EW} = (\tau_\mathrm{rec}/\tau_\mathrm{EW})^{-1/3}$~ \cite{Kahniashvili_Brandenburg_Tevzadze_2016, Brandenburg_Durrer_Huang_Kahniashvili_Mandal_Mukohyama_2020, Brandenburg:2021bfx}. 
Furthermore, using  Eq.~\eqref{eq:etaVSlambda}, we evolve the present-day values of $B_\mathrm{rms}^\mathrm{ex} = g'B_\mathrm{Y,rms}^\mathrm{ex}/e$ and $\lambda_B$ according to the scaling laws $B_\mathrm{rec}/B_\mathrm{EW} = (\tau_\mathrm{rec}/\tau_\mathrm{EW})^{-1/3}$ and $\lambda_\mathrm{rec}/\lambda_\mathrm{EW} = (\tau_\mathrm{rec}/\tau_\mathrm{EW})^{2/3}$~ \cite{Kahniashvili_Brandenburg_Tevzadze_2016, Brandenburg_Durrer_Huang_Kahniashvili_Mandal_Mukohyama_2020, Brandenburg:2021bfx}. {
To produce the correct baryon asymmetry,  the strength of the MF is required to $B_{\mathrm{rms}}^{\mathrm{ex}}\in[1.8\times10^{-16}, 5.3\times10^{-14}]$ Gauss for the correlation length $\lambda_B\in[2.4\times10^{-4},9.6\times10^{-4}]$ Mpc. Finally, using Eq. \eqref{eq:etaVSlambda} we can get the parameter space capable of achieving the correct baryon asymmetry, which is indicated by a blue band in Fig. \ref{fig:blazer}. } A significant portion of this parameter space is consistent with observational constraints. Therefore, the first-order electroweak PT in the presence of a hyperMF provides a viable mechanism for explaining the origin of matter-antimatter asymmetry.

%The evolution of its correlation length and strength is shown by the green line in Fig.~\ref{fig:blazer}.

%Following Eq.~\ref{eq:etaVSlambda}, the evolution of the MFs that can obtain the correct baryon asymmetry is shown by the yellow line ($\lambda_B\sim R_0$) and purple line ($\lambda_B\sim R_*$) in Fig. \ref{fig:blazer}.

\noindent{\it Conclusions.-}\label{sec:Con}
In this Letter, we numerically investigated the first-order electroweak PT in the background of a hyperMF. We found that a helical hyperMF can serve as a source of baryon asymmetry. We established the relationship between baryon asymmetry and the hyperMF, demonstrating that to achieve the correct matter-antimatter asymmetry, the present-day strength of a homogeneous background MF should be approximately $4.6\times10^{-18}$ Gauss. For a spectrum-distributed MF, the required field strength is $10^{-16}\sim10^{-14}$ Gauss with a correlation length of $10^{-4}\sim10^{-3}$ Mpc, which can be probed by Fermi-LAT gamma-ray observations.   

A key distinction of our work from previous studies, such as Refs.~\cite{Kamada:2016eeb, Kamada:2016cnb}, lies in the mechanism of baryon asymmetry generation. In our scenario, the MF produced during the first-order electroweak PT {together with the preexisting helical hyperMF} ensures the time variation of {the hypermagnetic helicity and Chern-Simons number}, which drives baryon asymmetry generation. In contrast, earlier studies considered baryon asymmetry arising from the decay of MF helicity before the SM cross-over, requiring an MF strength of around $10^{-17}$ Gauss at a correlation length of $10^{-3}$ pc. 

{
For the case of a homogeneous hyperMF, we further observed that the Higgs field forms a hexagonally arranged vortex when $g'B_Y^\mathrm{ex}/m_W^2\gtrsim 3$, regardless of whether the MF is helical or non-helical. 
This behavior aligns with the Ambj\o rn-Olesen condensation phenomenon~\cite{Nielsen:1978rm, PhysRevD.45.3833, SALAM1975203, LINDE1976435} predicted within the SM framework. The absence of this effect in Ref.~\cite{Kajantie:1998rz} for the first-order electroweak PT may be attributed to significant fluctuations in their configurations. Recent numerical confirmation of this phenomenon in the SM cross-over case can be found in Ref.~\cite{Chernodub_2023}. 
For hyperMFs with finite correlation lengths, the randomness of the field configuration prevents the observation of Ambj\o rn-Olesen condensation. Further details on the PT dynamics and Ambj\o rn-Olesen condensation in the background of the hyperMF can be found in the accompanied article~\cite{Di:2025ncl}. Furthermore, we did not observe the formation of electroweak strings as proposed in Refs.~\cite{PhysRevLett.101.171302, PhysRevLett.87.251302, Vachaspati:1994ng, Barriola:1994ez}, as our simulations did not impose any prior ansatz on the gauge fields, unlike the methodologies employed in those studies.
 }

We hope that our study provides valuable insights for future observations of cosmic MFs and helps uncover new physics beyond the SM. It is important to note that our simulations do not account for the expansion of the universe and the effects of relativistic fluids, both could influence the dynamics of the first-order electroweak PT. These effects warrant further investigation by the community. In addition, the gravitational waves generated by PTs in the presence of background MFs may differ from those without preexisting magnetic fields, offering another promising direction for future research. 

{The simulation strategy developed in this Letter can be directly generalized to study the topological defects evolution in the background of the primordial MF, such as cosmic string and domain wall that are considered as possible cosmic origins of the stochastic gravitational wave background excess the observed in the dataset of NANOGrav~\cite{Bian:2023dnv,NANOGrav:2023hvm}. }

%\section*{Acknowledgements}

 L.B. is supported by the National Key Research and Development Program of China under Grant No. 2021YFC2203004, and by the National Natural Science Foundation of China (NSFC) under Grants Nos. 12322505, 12347101. L.B. also acknowledges Chongqing Talents: Exceptional Young Talents Project No. cstc2024ycjh-bgzxm0020 and Chongqing Natural Science Foundation under Grant No. CSTB2024NSCQ-JQX0022. R.G.C. is supported by the National Key Research and Development Program of China Grants No. 2020YFC2201502 and No. 2021YFA0718304 and by National Natural Science Foundation of China Grants No.  12235019. The numerical calculations in this study were carried out on the ORISE Supercomputer.

\bibliography{reference}

\onecolumngrid
\begin{center}
  \textbf{\large Supplemental Material}\\[.2cm]
\end{center}

In the supplementary materials, we first detail the continuous and discrete equations of motion used in the simulation, and then provide the conditions for the initial Higgs field and the gauge fields. 

\section{Equations of Motion on Lattice}\label{App:EOM}
Under the temporal gauge $W_0=Y_0=Y^\mathrm{ex}_0 = 0$, the action on the lattice is
\begin{eqnarray}
    \nonumber
    S &=& \sum_{x,t} \Delta t \Delta x^3 \Bigg\{ \big(D_0\Phi \big)^\dagger \big(D_0\Phi \big) 
    - \sum_i \big(D_i\Phi \big)^\dagger \big(D_i\Phi \big) 
    -\mathcal{V}(\Phi) + \Big(\frac{2}{g\Delta t \Delta x}\Big)^2\sum_i \Big(1-\frac{1}{2}\text{Tr}~U_{0i}\Big) \\ 
    &&+\Big(\frac{2}{g'\Delta t \Delta x}\Big)^2 \sum_i \Big(1-\text{Re} ~V_{0i}\Big) 
    -\frac{2}{g^2 \Delta x^4} \sum_{i,j} \Big(1-\frac{1}{2} \text{Tr}~U_{ij}\Big) 
        - \frac{2}{g'^2\Delta x^4} \sum_{i,j} \Big(1-\text{Re}~V_{ij}\Big) \Bigg\} \notag \\
    &&+\Big(\frac{2}{g'\Delta t \Delta x}\Big)^2 \sum_i \Big(\mathrm{Im}~V_{0i}\Big) \Big(\mathrm{Im}~V_{\mathrm{ex}0i}\Big) 
    -\frac{2}{g'^2\Delta x^4} \sum_{i,j} \Big(\mathrm{Im}~V_{ij}\Big) \Big(\mathrm{Im}~V_{\mathrm{ex}ij}\Big).
    \label{actionlattice}
\end{eqnarray}
The Higgs field $\Phi$ is defined on the lattice sites while the gauge fields are defined on the links between two adjacent sites by link fields: 
\begin{align}
    U_i(t,x) &= 
    \exp\left[-\ii g\Delta x\frac{\sigma^a}{2} W_i^a(t,x)\right] \;,\\
    V_i(t,x) &= 
    \exp\left[-\ii g'\Delta x \,\frac12Y_i(t,x)\right] \;,\\
    V^\mathrm{ex}_i(t,x) &= 
    \exp\left[-\ii g'\Delta x \,\frac12Y^\mathrm{ex}_i(t,x)\right]\;,\label{eq:Vex}
\end{align}
Higgs field and gauge fields are defined on the exact time step, and their canonical momenta $\Pi,\ E,\ F$ are defined on the halfway between the time steps. 

We use $x+i$ to denote a space step forward towards the $i$ direction and $x-i$ 
to denote a space step backward. 
The link fields, $U_i(t,x),\ V_i(t,x),\ V^\text{ex}_i(t,x)$, parallel transport
the Higgs field $\Phi$ that located at site $x+i$ back to site $x$; and their hermitian conjugate
parallel transport the Higgs field $\Phi$ that located at site $x$ to site $x+i$. 
Therefore, Using the leapfrog algorithm, the covariant derivative of $\Phi(t,x)$ is
\begin{align*}
	D_i \Phi &= \frac{1}{\Delta x} \big[ U_i(t,x) V_i(t,x) V_{\mathrm{ex}i}(t,x) \Phi(t,x+i) - \Phi(t,x)\big]\;, \\
	D_0 \Phi &= \frac{1}{\Delta t} \big[ U_0(t,x) V_0(t,x) V_{\mathrm{ex}0}(t,x) \Phi(t+\Delta t,x) -\Phi(t,x) \big] \\
    &= \frac{1}{\Delta t}\big[ \Phi(t+\Delta t,x) -\Phi(t,x) \big]\;.
\end{align*}
The plaquette fields with all space indices are  
\begin{align*}
    U_{ij}(t,x) &= U_j(t,x) U_i(t,x+j) U_j^\dagger (t,x+i) U_i^\dagger (t,x)\;, \\
    V_{ij}(t,x) &= V_j(t,x) V_i(t,x+j) V_j^\dagger (t,x+i) V_i^\dagger (t,x) \;,\\
    V_{\mathrm{ex}ij}(t,x) &= V_{\mathrm{ex}j}(t,x) V_{\mathrm{ex}i}(t,x+j) V_{\mathrm{ex}j}^\dagger (t,x+i) V_{\mathrm{ex}i}^\dagger (t,x)\;.
\end{align*}
The plaquette fields with one time indices are
\begin{align*}
    U_{0i}(t,x) &= U_i(t,x)U_0(t+\Delta t,x)U_i^\dagger(t+\Delta t,x)U_0^\dagger(t,x) 
    = U_i(t,x)U_i^\dagger(t+\Delta t,x)\;, \\
    V_{0i}(t,x) &= V_i(t,x)V_0(t+\Delta t,x)V_i^\dagger(t+\Delta t,x)V_0^\dagger(t,x) 
    = V_i(t,x)V_i^\dagger(t+\Delta t,x)\;, \\
    V_{\mathrm{ex}0i}(t,x) &= V_{\mathrm{ex}i}(t,x)V_{\mathrm{ex}0}(t+\Delta t,x)
    V_{\mathrm{ex}i}^\dagger(t+\Delta t,x)V_{\mathrm{ex}0}^\dagger(t,x)
    = V_{\mathrm{ex}i}(t,x)V_{\mathrm{ex}i}^\dagger(t+\Delta t,x)\;.
\end{align*}

\section{Initialization}\label{App:init}

Before the PT occurs, we believe that the Higgs field is in equilibrium. Therefore, we can use the thermal spectrum to describe the amplitude and momentum distribution of each scalar component $\Phi_i$ of the Higgs doublet $\Phi$ and each scalar component $\Pi_i$ of the conjugate momentum field $\Pi$ in momentum space
\begin{align}
    \mathcal{P}_{\Phi_i}(k) = \frac{1}{\omega_k} \frac{1}{e^{{\omega_k}/T} - 1}, \qquad
    \mathcal{P}_{\Pi_i}(k) = \frac{\omega_k}{e^{{\omega_k}/T} - 1},
\end{align}
and
\begin{align}
    \Phi = \frac{1}{\sqrt{2}}\begin{pmatrix}
        \Phi_1 + \ii\Phi_2 \\ \Phi_3 + \ii\Phi_4
    \end{pmatrix}, \qquad
    \Pi = \frac{1}{\sqrt{2}}\begin{pmatrix}
        \Pi_1 + \ii\Pi_2 \\ \Pi_3 + \ii\Pi_4
    \end{pmatrix},
\end{align}
where $(e^{{\omega_k}/T} - 1)^{-1}$ is the occupation number of the Bose-Einstein distribution, $\omega_k = \sqrt{k^2 + m^2_\mathrm{eff}}$ is the physical frequency, $k$ and $m_\mathrm{eff}$ are physical momentum and effective mass of the Higgs field.

In the continuum, we have
\begin{align}
    \langle\Phi_i(\boldsymbol{k})\Phi_j(\boldsymbol{k}')\rangle &= (2\pi)^3\mathcal{P}_{\Phi_i}(k)\delta(\boldsymbol{k}-\boldsymbol{k}')\delta_{ij}, \label{eq:phithermal}\\
    \langle\Pi_i(\boldsymbol{k})\Pi_j(\boldsymbol{k}')\rangle &= (2\pi)^3\mathcal{P}_{\Pi_i}(k)\delta(\boldsymbol{k}-\boldsymbol{k}')\delta_{ij}, \\
    \langle\Phi_i(\boldsymbol{k})\Pi_j(\boldsymbol{k}')\rangle &= 0.
\end{align}
Converting it into a discrete form on the lattice, we get
\begin{align}
    \langle|\Phi_i(\boldsymbol{k})|^2\rangle = \left(\frac{N}{\Delta x}\right)^3\mathcal{P}_{\Phi_i}(k), \qquad
     \langle\Phi_i(\boldsymbol{k})\rangle = 0, \\
     \langle|\Pi_i(\boldsymbol{k})|^2\rangle = \left(\frac{N}{ \Delta x}\right)^3\mathcal{P}_{\Pi_i}(k), \qquad
     \langle\Pi_i(\boldsymbol{k})\rangle = 0,
\end{align}
where $N$ denotes the number of points per side and $\Delta x$ is the physical lattice spacing. $\Phi_i(\boldsymbol{k})$ and $\Pi_i(\boldsymbol{k})$ satisfy the Gaussian distribution in the momentum space that varies from point to point, and contains all modes from infrared truncation $k_\mathrm{IR} = 2\pi/(N\Delta x)$ to the ultraviolet cutoff $k_\mathrm{CutOff}$. 

For the gauge fields, we set their initial values to $0$, then the values of the link fields are $1$ for U(1) or $1_{2\times2}$ for SU(2). However, to satisfy the Gaussian constraint
\begin{align}
	&\partial_0 \partial_j Y_j -g'\text{Im}\big[ \Phi^\dagger \partial_0 \Phi \big] = 0\;, \notag \\ 
	& \partial_0 \partial_j W^a_j + g \epsilon^{abc} W^b_j \partial_0 W^c_j 
        -g\text{Im} \big[ \Phi^\dagger \sigma^a \partial_0 \Phi \big] =0,
\label{gauss-2}
\end{align}
we must assign an initial value to the conjugate momentum field of the gauge field as follows.
\begin{align}
    &\frac{1}{\Delta x}\sum_i
        \mathrm{Im}\big[E_i(t,x)-E_i(t,x-i)\big] = g'\mathrm{Im}[\Pi^\dagger(t,x)\Phi(t,x)], \label{U1gaussinit}\\
    &\frac{1}{\Delta x}\mathrm{Tr}
            \sum_i\ii\sigma^m\big[F_i(t,x)-U^\dagger_i(t,x-i)F_i(t,x-i)U_i(t,x-i)\big] = g\mathrm{Re}[\Pi^\dagger(t,x)\ii\sigma^a\Phi(t,x)].
\end{align}

\section{Hypermagnetic Field on Lattice}\label{App:Mag}
The MF is not well defined before the electroweak PT occurs, we therefore use an extra external hyperMF $\boldsymbol{B}^\mathrm{ex}_Y = \nabla \times \boldsymbol{Y}^\mathrm{ex}$ 
instead of MF 
$\boldsymbol{B}^\mathrm{ex} = \nabla \times \boldsymbol{A}^\mathrm{ex}$ \cite{Chernodub_2023}, where
$\boldsymbol{Y}^\mathrm{ex}$ is the space part of U(1) gauge field $Y^\mu_\mathrm{ex} = (Y^0_\mathrm{ex},\ \boldsymbol{Y}_\mathrm{ex})$ and
$\boldsymbol{A}_\mathrm{ex}$ is the space part of electroMF $A^\mu_\mathrm{ex} = (A^0_\mathrm{ex},\ \boldsymbol{A_\mathrm{ex}})$. 
In the broken phase, the relation between the two fields is 
$g'\boldsymbol{B}_Y^\mathrm{ex} = e\boldsymbol{B}^\mathrm{ex}$, here
$e=g\sin\theta_w$ is the electric charge. 

To achieve a helical homogeneous hyperMF along the $z$ direction, 
one can easily find that the U(1) field has the following form: 
\begin{align}
    Y^{\mathrm{ex}\mu}(t,x,y,z) = (0,\ 0,\ xB_Y^\mathrm{ex},\ h_\mathrm{factor}LB_Y^\mathrm{ex}), 
\end{align}
since
\begin{align}
    \nabla \times \boldsymbol{Y}^\mathrm{ex}
    = \begin{vmatrix}
        \hat x & \hat y & \hat z \\
        \partial_x & \partial_y & \partial_z \\
        0 & xB_Y^\mathrm{ex} & h_\mathrm{factor}LB_Y^\mathrm{ex}
    \end{vmatrix}
    = \partial_x(xB_Y^\mathrm{ex})\hat z = B_Y^\mathrm{ex}\hat z, 
\end{align}
where $B_Y^\mathrm{ex}= \frac{2}{g'} 2\pi N_BL^{-2} \label{qc},\ N_B\in \mathbb{Z}$~\cite{Bali_2012} is a constant.

For the hyperMF having the form of 
\begin{align}
    \Tilde{B}_{Yi}^\mathrm{ex}(\boldsymbol{k}) = B_\mathrm{ini}\left(\delta_{ij}-\hat{k}_i\hat{k}_j-\ii\sigma_\mathrm{M}\varepsilon_{ijl}\hat{k}_l\right)g_j(\boldsymbol{k})k^n, \label{eq:specB}
\end{align}
in Fourier space, we can first verify that 
\begin{align}
    \Tilde{Y}^\mathrm{ex}_l(\boldsymbol{k}) = \ii B_\mathrm{ini}\left(\varepsilon_{lmn}\hat{k}_m-\ii\sigma_\mathrm{M}\delta_{ln}\right)g_n(\boldsymbol{k})k^{n-1} \label{eq:specY}
\end{align}
is its corresponding vector potential.
Given that $\boldsymbol{B}^\mathrm{ex}(\boldsymbol{x})=\nabla\times\boldsymbol{Y}^\mathrm{ex}(\boldsymbol{x})$, doing an inverse Fourier transform gives 
\begin{align}
    \int \frac{\mathrm{d}^3\boldsymbol{k}}{(2\pi)^3}\,\Tilde{\boldsymbol{B}}^\mathrm{ex}(\boldsymbol{k})e^{\ii\boldsymbol{k\cdot x}} &= \nabla\times\int \frac{\mathrm{d}^3\boldsymbol{k}}{(2\pi)^3}\,\Tilde{\boldsymbol{Y}}^\mathrm{ex}(\boldsymbol{k})e^{\ii\boldsymbol{k\cdot x}} \\
    &= \int \frac{\mathrm{d}^3\boldsymbol{k}}{(2\pi)^3}\,\ii\boldsymbol{k}\times\Tilde{\boldsymbol{Y}}^\mathrm{ex}(\boldsymbol{k})e^{\ii\boldsymbol{k\cdot x}}.
\end{align}
From this we get $\boldsymbol{\Tilde{B}}^\mathrm{ex}(\boldsymbol{k})=\ii\boldsymbol{k}\times\boldsymbol{\Tilde{Y}}^\mathrm{ex}(\boldsymbol{k})$ or $\Tilde{B}^\mathrm{ex}_i(\boldsymbol{k})=\ii\varepsilon_{ijl}k_j\Tilde{Y}^\mathrm{ex}_l(\boldsymbol{k})$. Then using \eqref{eq:specY}, we have
\begin{align}
    \ii\varepsilon_{ijl}k_j\Tilde{Y}^\mathrm{ex}_l(\boldsymbol{k}) &= -\varepsilon_{ijl}k_j B_\mathrm{ini}\left(\varepsilon_{lmn}\hat{k}_m-\ii\sigma_\mathrm{M}\delta_{ln}\right)g_n(\boldsymbol{k})k^{n-1} \notag\\
    &= -B_\mathrm{ini}\left[(\delta_{im}\delta_{jn}-\delta_{in}\delta_{jm})\hat{k}_m\hat{k}_j-\ii\sigma_\mathrm{M}\varepsilon_{ijn}\hat{k}_j\right]g_n(\boldsymbol{k})k^{n} \notag\\
    &= B_\mathrm{ini}\left(-\hat{k}_i\hat{k}_n+\delta_{in}+\ii\sigma_\mathrm{M}\varepsilon_{ijn}\hat{k}_j\right)g_n(\boldsymbol{k})k^{n} \notag\\
    &= B_\mathrm{ini}\left(\delta_{ij}-\hat{k}_i\hat{k}_j-\ii\sigma_\mathrm{M}\varepsilon_{ijl}\hat{k}_l\right)g_j(\boldsymbol{k})k^{n},
\end{align}
which is consistent with \eqref{eq:specB}. 

To obtain the external hyperMF of \eqref{eq:specB}, one should follow the steps below:
\begin{itemize}
    \item Step 1: Generates a $\delta$-correlated Gaussian distributed random vector field $\boldsymbol{g}(\boldsymbol{x})$ in coordinate space.
    \item Step 2: Perform Fourier transform on $\boldsymbol{g}(\boldsymbol{x})$ and get $\boldsymbol{g}(\boldsymbol{k}) = \int\mathrm{d}^3\boldsymbol{x}\, \boldsymbol{g}(\boldsymbol{x})e^{-\ii\boldsymbol{k\cdot x}}$.
    \item Step 3: According to \eqref{eq:specY}, we can get $\boldsymbol{\Tilde{Y}}^\mathrm{ex}(\boldsymbol{k})$.
    \item Step 4: Perform an inverse Fourier transform on $\boldsymbol{\Tilde{Y}}^\mathrm{ex}(\boldsymbol{k})$ to obtain $\boldsymbol{Y}^\mathrm{ex}(\boldsymbol{x}) = \int \mathrm{d}^3\boldsymbol{k}\,\Tilde{\boldsymbol{Y}}^\mathrm{ex}(\boldsymbol{k})e^{\ii\boldsymbol{k\cdot x}}/(2\pi)^3$.
    \item Step 5: Using \eqref{eq:Vex}, we can finally obtain the distribution of the link field $V^\mathrm{ex}_i$ on the littice.
\end{itemize}

\end{document}